\documentclass[10pt,preprint,a4paper]{aastex}

\usepackage{amsmath}                
\usepackage{amsfonts}               
\usepackage{amssymb}                
\usepackage{epsfig}                 

\newcommand{\s}{{~\rm s}}
\newcommand{\km}{{~\rm km}}

\newcommand{\yr}{{~\rm yr}}

\def \astrobj#1{#1}

\title{FINAL COMMON ENVELOPE EJECTION BY MIGRATION AND JETS}

\author{Noam Soker\altaffilmark{1}}

\altaffiltext{1}{Department of Physics, Technion -- Israel Institute of
Technology, Haifa 32000 Israel; soker@physics.technion.ac.il}

\begin{document}

\begin{abstract}
I summarize recent analytical and numerical studies of the common envelope (CE) process and
suggest to replace the commonly used $\alpha_{\rm CE}$-prescription for the CE ejection by
a prescription based on final migration and jets launched by the companion or the core of the giant stellar primary.
In the migration process the core-companion binary systems is surrounded by a highly oblate (flatten) envelope, a thick circumbinary disk,
formed by the large angular momentum transferred from the core-companion system to the envelope.
I then show that the energy that can be released by an accreting main sequence companion can surpass the mutual
gravitational energy of the core and the companion. An efficient channel to leash the accretion energy to
expel the CE is through jets operating via a feedback mechanism (JFM).
\end{abstract}



\section{INTRODUCTION}
\label{sec:introduction}

During the common envelope (CE) phase of a binary system the orbital separation between the core of the large star and the
smaller secondary star decreases due to gravitational drag and tidal interaction
(e.g., \citealt{Paczynski1976, vandenHeuvel1976, IbenLivio1993, TaamSandquist2000, Podsiadlowski2001, Webbink2008, TaamRicker2010, RickerTaam2012, Ivanovaetal2013}).
The exact process that determines the core-secondary final orbital separation is one of the major unsolved questions of the CE process.
Another related open question is the duration of the final phase of the CE evolution, being days to several weeks (e.g., \citealt{RasioLivio1996}; \citealt{LivioSoker1988}), or
months (e.g., \citealt{SandquistTaam1998, DeMarco2003, DeMarco2009, Passy2011, RickerTaam2012}).

In the commonly used $\alpha_{\rm CE}$-prescription the gravitational energy released by the spiraling-in binary system, $E_G$, is
equated to the envelope binding energy (e.g., \citealt{Webbink1984, TaurisDewi2004, Ivanovaetal2013}), $E_{\rm bind}$,
with an efficiency of $\alpha_{\rm CE}$: $\alpha_{\rm CE} E_G = E_{\rm bind}$.
Many researchers include the internal energy of the envelope (e.g. \citealt{Han1994, Maxted2002, Zorotovic2010, XuLi2010, Davis2011, Rebassa-Mansergas2012}),
or the enthalpy \citep{IvanovaChaichenets2011}, in the energy-balance equation

In the $\alpha_{\rm CE}$-prescription the CE is ejected in a uniform manner and there is no separation between envelope parts.
However, numerical simulations have shown the separation of the CE to ejected and bound segments, e.g., \cite{Lombardi2006}.
\cite{SandquistTaam1998}, as another example, found that only about a quarter of the CE is ejected; the rest of the envelope remains bound to one or both of the interacting stars.
They also obtained a differentially rotating thick disk or torus at intermediate stages of the CE evolution.
\cite{Soker1992} (see also \citealt{Soker2004}) analytically obtained a similar thick disk structure.
In these cases a rapid merging is expected as well, unless an extra energy source is applied, such as accretion onto the secondary star.
\cite{Passy2011} found in their simulations that when the envelope is lifted away from the binary,
$\gtrsim 80$ per cent of the envelope remains bound to the binary system.
\cite{DeMarco2011} suggested that envelope material still bound to the binary system at the end of the CE
falls back, forms a circumbinary disk, and influences further binary evolution.
\cite{RickerTaam2012} find that during the rapid inspiral phase (the plunge-in phase; see \citealt{Ivanovaetal2013})
only $\sim 25 \%$ of the energy released by the spiraling-in process goes toward ejection of the envelope.
\cite{Passyetal2012} made a detailed study of the CE evolution and found that not only most of the CE stays bound ($\ga 90 \%$),
but the final orbital separations in their simulations are much larger than those observed.

The separation of the CE to ejected and bound parts was shown also analytically by \citet{KashiSoker2011}.
They argue that the $\sim 1-10 \%$ of the ejected envelope that remains bound falls-back and forms a circumbinary disk
around the post-CE binary system.
The interaction of the circumbinary disk with the binary system further reduces the orbital separation, leading in many cases
to a merger at the end of the CE phase or a short time after.
A different mechanism for merger at the termination of the CE phase was suggested by \cite{IvanovaChaichenets2011}.

In a recent paper \citep{Soker2013} I put another basic assumption of the $\alpha_{\rm CE}$-prescription into question.
The assumption is that most of the core-secondary gravitational energy, that is liberated in the final
spiraling-in process, can be used efficiently to unbind the envelope.
In that paper I found that when the orbital separation decreases to $\sim 10$ times the final orbital separation predicted by
the $\alpha_{\rm CE}$-prescription, the companion has not enough mass in its vicinity to carry away its angular momentum;
see figure 18 in \cite{Passyetal2012} for a numerical demonstration of this effect.
Instead, the binary system interacts gravitationally with a rapidly-rotating flat envelope, as is found in the numerical
simulations cited above. This situation resembles that of planet-migration in protoplanetary disks.
The envelope convection of the giant star carries energy and angular momentum outward.

Both analytical calculations and numerical simulations show that merger might be a common outcome of the CE evolution.
Actually, merger might be too common if no extra energy source to remove the envelope is applied.
In this paper I put forward the idea that jets launched by the secondary star act via a feedback mechanism to
remove the rest of the envelope.
The idea of using jets launched by the secondary star to help removing the CE was proposed in the past, but under specific conditions.
In a previous paper \citep{Soker2004} I discussed the removal of the CE by jets launched by a
{{{ { sufficiently compact secondary star, such that an accretion disk is formed via the Bondi-Hoyle-Lyttleton (BHL) accretion process from the envelope.
I showed that such a companion cannot be a main sequence (MS) star, but rather must be a} }}} neutron star (NS) or a white dwarf (WD).
\cite{Chevalier2012} then explored the possibility that the mass loss prior to an explosion of a core-NS merger process is
driven by a CE evolution of a NS (or a BH) in the envelope of a massive star.
This explosion process was further developed by \cite{Papishetal2013}.
In the present study I concentrate on main sequence (MS) stellar companions, that are the more common types of secondaries
in CE systems.
{{{  {The major difference from previous studies, including that of \cite{Soker2004}, is that here the main accretion process is not the BHL process, but rather an accretion from
a highly distorted oblate envelope residing outside the secondary orbit. Such a process can take place only when the secondary is deep inside the envelope.} }}}
General properties of the proposed mechanism are discussed in section \ref{sec:jets}, and some quantitative properties of
the flow are studied in section \ref{sec:energy}. A summary is in section \ref{sec:summary}.

\section{PHASES OF JET ACTIVITY}
\label{sec:jets}

\begin{figure}[!htb]
\centering
\includegraphics[width=0.85\textwidth]{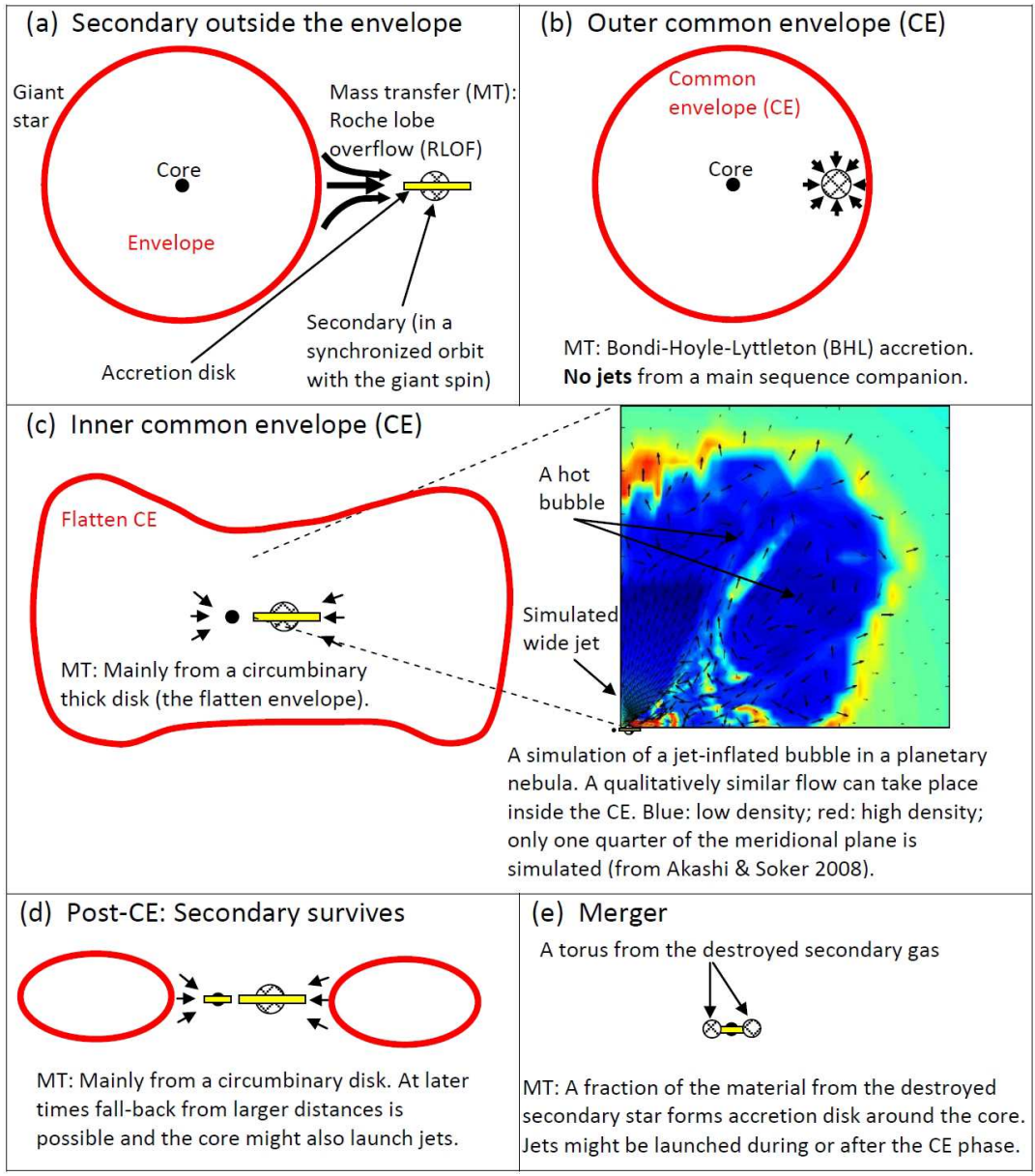}
\caption{Schematic drawing (not to scale) of the phases considered here that are summarize in Table 1.
The study concentrates on a main sequence (MS) companion interacting with an evolved AGB star.
The image on the right side of panel c is a numerical simulation of a jet-inflated bubble in a PN taken from \cite{AkashiSoker2008}.
I suggest that a qualitatively similar flow takes place in the final CE evolution: bubbles inflated by jets launched by the companion
remove a significant part of the leftover envelope. This is the negative-feedback cycle.
Vortices induced by the bubbles channel some envelope mass toward the plane to feed the accretion disk. This is the positive-feedback cycle.
}
\label{figure:fig1}
\end{figure}

I examine the conditions for accretion disk formation, and assume that a massive accretion disk launches jets.
For the formation of an accretion disk the specific angular momentum of the accreted mass must be sufficiently large.
This in turn requires a highly asymmetric accretion flow.  In Table 1 I list the phases when an accretion
disk might be formed. These are drawn schematically in Fig. \ref{figure:fig1}.
\begin{table}[!htb]
\label{Tab:Table1}
Table 1: Jet activity phases
\newline
\bigskip
\begin{tabular}{|l|l|l|l|l|}
\hline
\small {Secondary star}&{Mass source}&{Role of Jets}&{Observational signatures}& {Comments} \\
\hline
\small (a) Outside envelope.   & RLOF         & Shaping the slow   & Bipolar symbiotic            &          \\
\small Mainly in               &              & giant wind.        & nebulae; Bipolar PNe         &           \\
\small synchronization.        &              &                    & with a narrow waist.         &           \\
\hline
\small (b) Outer CE.           & Bondi-Hoyle  & No jets from       &                              & Possible jets         \\
\small Spirals-in.             & -Lyttleton   & a main sequence    &                              & from NS and WD        \\
\small                         & accretion.   & secondary.         &                              & secondaries.          \\
\hline
\small (c) Inner CE.           & Circumbinary & Removing and       & Elliptical PNe.              & A feedback          \\
\small Migrates-in due to      & thick disk   & accelerating       &                              &  process.        \\
\small circumbinary disk.      & (flatten CE).& the CE.            &                              &           \\
\hline
\small (d) Post CE.            & Circumbinary & Shaping the nebula;& Elliptical PNe+ansae;        & Jets might be          \\
\small Residual migration.     & thick disk or& Forming hot        & Mildly bipolar PNe.          & lunched also           \\
\small                         & fall-back gas.& bubbles in        & Diffuse X-ray in             & from the core.          \\
\small                         &              &  elliptical PNe    & elliptical PNe.              &           \\
\hline
\small (e) Merger (during or   & Destroyed    & Shaping the nebula;& Elliptical PNe+ansae;        & Jets might            \\
\small after CE ejection).     & secondary or & Forming hot        & Mildly bipolar PNe.          & be highly     \\
\small Core launches jets.     & fall-back gas.& bubbles in         & Diffuse X-ray in             & collimated.           \\
\small                         &              &  elliptical PNe.    & elliptical PNe.              &           \\
\hline
\end{tabular}
 \footnotesize
\newline
The table refers to a late AGB primary star and a main sequence (MS) secondary (companion) star. Many items are relevant
to other types of binary systems. Schematic illustrations of the different phases are drawn in Fig. \ref{figure:fig1}.
`Ansae' stand for two opposite small bullets, one at each side of an elliptical PN, that generally move faster than the rest of the nebula.
\end{table}

(a) When the secondary is outside and close to the giant envelope a Roche lobe overflow (RLOF) process,
or a similar one if the orbit is not synchronized with the giant's spin, supplies mass with sufficiently
high specific angular momentum to form an accretion disk. The companion can be a MS star, a WD, a NS, or a BH.
In the case of a WD companion a bipolar symbiotic nebula can be formed.
Later on, whether the companion enters or not a CE, a bipolar PN with narrow waist can be formed.

(b) I consider the formation of a CE due to further evolution of the giant and/or angular momentum loss in the wind.
As long as the accretion process onto the secondary star during the CE evolution is the Bondi-Hoyle-Lyttleton (BHL) accretion process, the
specific angular momentum of the accreted mass is low, and no accretion disk will be formed around a MS star; disks
might be formed around WD or NS companions \citep{Soker2004}. However, a WD cannot accrete at a high rate, as the hydrogen is ignited
and the outer layers of the accreting WD are inflated to prevent further accretion.

(c) When the companion is deep inside the envelope, a flatten envelope is formed \citep{RickerTaam2012, Passyetal2012}.
The flatten envelope might have two effects on the core-secondary binary system.
First, a tidal interaction of the binary system with the flatten envelope residing outside the secondary orbit, which is a thick circumbinary disk,
can lead to reduction of the orbital separation \citep{KashiSoker2011}, in what is termed here a migration process \citep{Soker2013}.

Second, the mass accreted from the circumbinary disk is likely to have sufficient specific angular momentum to form an accretion disk around the
secondary star, {{{ {even if it is a MS star.} }}}
If a disk is formed, it is likely to launch jets, as shown schematically in panel c of Fig. \ref{figure:fig1},
where a flow pattern from a different numerical setting is presented.
The simulation presented is of a wide jet launched inside a spherical nebula, to mimic the formation of a bipolar PNe \citep{AkashiSoker2008}.
The simulation is of the entire volume, but a cylindrical symmetry is assumed, hence the numerical grid is 2D.
Symmetry is assumed around the jet's axis (vertical axis in the figure), and a mirror symmetry is assumed about the orbital plane (horizontal line in the figure).
Only one quarter of the meridional plane is simulated and presented.
{{{  {The simulation is for a different setting than the CE jets studied here in that the companion is outside the envelope. It is presented only to demonstrate the formation of a hot bubble.
It is not intended to present the the formation of the accretion disk.} }}}
A qualitatively similar flow, I suggest, can take place when jets are launched by the secondary star inside the envelope.
As discussed in the next section, for example, the orbital motion of the secondary star leads to wide jets.

{{{ {I emphasize that the formation of an accretion disk in phase c is a new ingredient. In \cite{Soker2004} I only studied the formation of an accretion disk
via the BHL accretion process from the envelope, where an accretion disk around a MS star cannot be formed.
In phase c the process is more like that in young stellar objects, where a circum-binary disk feeds the stars, and an accretion disk can be formed around a MS star.
When the binary system is deep in the envelope, the highly distorted circum-binary oblate envelope contains most of the angular momentum of the system.
Due to tidal interaction and friction in the rotating envelope, I suggest that the envelope feeds the binary system with gas possessing sufficiently high specific angular momentum.
This claim will have to be studied with 3D numerical simulations. The numerical study of this feeding process from a rotating oblate envelope is more complicated
than the 3D numerical studies of the CE phase that have been conducted till now (and much more complicated than the feeding from a geometrically thin accretion disk),
but are highly encouraged. } }}}

The removal of the envelope with the inflation of hot low-density bubbles operates through a feedback cycle composed of a positive part and a negative part.
The positive part is the processes by which the bubbles push material toward the equatorial plane \citep{AkashiSoker2008}, that can further
supply mass to the secondary and amplify the jets power. The negative part is simply the removal of envelope mass, hence reducing the accretion rate.

(d) After the removal of the envelope a residual circumbinary disk might be left. Its life can be prolonged by fall back of nebular gas.
Post CE jets (PCEJ) might be launched. Due to the orbital motion they will not be well collimated, and might show signs of precessions.
At later times, when the leftover envelope mass on the core is removed and shrinks, the core itself might accrete mass through a disk, and lunch jets.
\cite{SokerLivio1994} considered the launching of jets before and after the CE phase, but not during the CE phase.
In the PCE phase they considered mass transfer from the secondary to the core and fall back gas, hence the core launches the jets.
Both in the final CE phase and in the post-CE phase mass can be transferred from the leftover envelope residing on the core to the secondary star.
This mass transfer is also likely to form an accretion disk. Namely, in the energy balance there is no need to include the envelope mass residing
within $\sim 0.5 R_\sun$ from the core.

(e) Finally, the secondary can merge with the core \citep{Soker2013} and forms and accretion disk. In \cite{Soker1996} I proposed this
mechanism for brown dwarf and massive planet companions. Here I extend it to low mass MS stars (low mass relative to the envelope mass).
As there is no orbital motion any more, the jets might be well collimated.

\section{AVAILABLE ENERGY}
\label{sec:energy}

If the jet feedback mechanism (JFM) is responsible for the removal of a significant part of the envelope, then the energy $E_{\rm jets}$
released by accreting mass $M_{\rm acc}$ onto the companion of mass $M_2$ and radius $ R_2$
{{{ {should be comparable or larger than the binding energy of the CE (at least the envelope residing above the final orbital separation).
I derive the required accreted mass onto the companion from this condition.
Instead of the explicit expression for the CE binding energy, I use the equivalent energy from the $\alpha_{\rm CE}$-prescription, $E_{\alpha_{\rm CE}}$.} }}}
A plausible scaling gives
\begin{equation}
E_{\rm jets} \simeq \frac {G M_2 M_{\rm acc}}{2R_2} =
\left( \frac{\alpha_{\rm CE}}{0.5} \right)^{-1}
\left( \frac{M_{\rm acc}}{0.1M_{\rm core}} \right)
\left( \frac{R_2}{0.2 a_{\rm final}} \right)^{-1}   E_{\alpha_{\rm CE}}.
\label{eq:ejet1}
\end{equation}
To be consistent with the definition of $E_{\rm jets}$ I take
\begin{equation}
E_{\alpha_{\rm CE}} = \frac {G M_{\rm core}{M_2}}{2a_{\rm final}}\alpha_{\rm CE},
\label{eq:alphace1}
\end{equation}
and $a_{\rm final}$ is the final core-companion orbital separation according to the $\alpha_{\rm CE}$-prescription.
I also assume that at these very high accretion rates most of the liberated accretion energy is channelled to outflow at about the escape
speed from the secondary star.

Not all the accretion energy will be channelled to the jets. On the other hand there are two other processes to consider.
(1) The spiral-in process already lifted part of the envelope. As well, the migration processes will release energy and will further
inflate the envelope. The binding energy of the inflated envelope is lower than its pre-CE value.
(2) The very inner envelope mass that is close to the core, within $\sim 0.5 R_\sun$, does not need to be removed.
It can be accreted onto the secondary star. This segment of the envelope has a high binding energy, which is removed from the energy budget in the
JFM for CE removal.

Over all, to equate the energy given by the classical $\alpha_{\rm CE}$-prescription, $E_{\rm jets} \simeq E_{\alpha_{\rm CE}}$,
the companion should accrete a mass of
\begin{equation}
M_{\rm acc} \simeq 0.06
\left( \frac{M_{\rm core}}{0.6 M_\sun} \right)
\left( \frac{R_2}{0.5 R_\sun} \right)
\left( \frac{a_{\rm final}}{2.5 R_\sun} \right)^{-1}
\left( \frac{\alpha_{\rm CE}}{0.5} \right)
M_\sun,
\label{eq:macc1}
\end{equation}
which is another presentation of equation (\ref{eq:ejet1}).
The parameters $\alpha_{\rm CE}$ and $a_{\rm final}$ appear here as $a_{\rm final}/\alpha_{\rm CE} = 5 R_\sun$,
which is compatible in general terms with the findings of \cite{DeMarco2011}.
If the mass in the jets is half the accreted mass, then the scaling in equation (\ref{eq:macc1}) requires, for example, that a mass of $0.03 M_\sun$ be lost in the jets
at a velocity of $\sim 620 \km \s^{-1}$ (similar to the solar wind speed).

{{{ {Very crudely, the migration phase lasts for $\sim 1 \yr$ \citep{KashiSoker2010}.
In the present scenario the accretion from the envelope outside the companion orbit starts when the envelope is highly distorted to an oblate shape.
The total accretion period is longer than the migration phase, but not by much. If accretion occurs intermittently, then the effective accretion time is
somewhat shorter.
Overall, the accretion period lasts for several months to more than a year, and the accretion rate onto the companion is
$\sim 0.05 -0.1 M_\odot \yr^{-1}$ for the above scaling. In one year the star accretes $\sim 0.1$ of its mass,
as might have been the case in the Great Eruption of \astrobj{Eta Carinae} \citep{KashiSoker2010}.
At these very high accretion rates it has been suggested that the mass ejected in the jets is a large fraction of the accreted mass, up to $\sim 0.5$, e.g.,
\cite{KashiSoker2010} in their modelling of the Great Eruption of \astrobj{Eta Carinae}.
The accretion period is $\sim 100 $ times the typical dynamical time of the accretion disk around a MS star (at several stellar radii),
and there is no unusual demands on the viscosity in the disk. } }}}

{{{ {The response of a MS star to such a high accretion rate sensitively depends on the energy content of the accreted mass (or its entropy).
On the inner boundary of an accretion disk touching the accreting star, the gravitational energy value is twice that of the kinetic energy. Namely, the accreted gas
already obeys the virial relation, implying that the accreting star does not need to radiate extra energy or heat much the accreted mass. The accreting star radius will not change much.
As well, the $\sim 0.06 M_\odot$ accreted mass will be mixed in the convective envelope of low mass MS stars.
In the envisioned scenario presented here the jets carry a large fraction of the accreted energy, such that the accreted gas might have a kinetic energy less than half the
value of the gravitational energy. For example, this can happen if some fraction of the energy in the boundary layer, where disk-kinetic energy is transferred to thermal energy, is removed by jets.
In such a case an accreting MS star might even shrink.} }}}

The ratio of the orbital velocity of the secondary to the jet velocity in the final CE phase is
\begin{equation}
\frac {v_{\rm orb2}}{v_{\rm jet}} =
0.25
\left( \frac{v_{\rm jet}}{650 \km \s^{-1}} \right)^{-1}
\left( \frac{M_{\rm core}+M_2}{1.0 M_\sun} \right)^{-1/2}
\left( \frac{M_{\rm core}}{0.6 M_\sun} \right)
\left( \frac{a_{\rm final}}{2.5 R_\sun} \right)^{-1/2}.
\label{eq:vorb1}
\end{equation}
The ratio ${v_{\rm orb2}}/{v_{\rm jet}} \sim 0.25 $ implies that over an orbit the jets will be launched on a wide angle, even if the jets
are narrow at ejection. If the jets have a half opening angle at source of $10^\circ$, for example, with the orbital velocity
given in equation (\ref{eq:vorb1}) the half opening angle over an orbit is $\sim 25 ^\circ$.
(At each moment the jets still have an opening angle of $10^\circ$, but due to the orbital motion at each orbital phase the jets will be bent by $\sim 15^\circ$
to a different direction.)
The jets can interact with a large fraction of the CE volume.
{{{ {Because of the orbital motion the jets' axis is constantly displaced and the jets encounter fresh envelope material.
The jets don't manage to penetrate through the envelope, and hence deposit their energy inside the envelope \citep{Soker2004, Papishetal2013}. } }}}

The inequality $({v_{\rm jet}}/v_{\rm orb2})^2 \ga 10$ implies that the shocked jet material will be hotter than the gas
temperature of the envelope near and outside the secondary orbit. Hence, low density hot bubbles will be formed.
The later evolution of these bubbles must be studied in more details, and with 3D numerical simulations.

\section{SUMMARY}
\label{sec:summary}

I proposed a scenario where the final removal of the common envelope (CE) in cases where a merger of the secondary with the core is avoided
is done by jets launched by the secondary star.
This process is suggested to replace the classical $\alpha_{\rm CE}$-prescription for the CE final ejection.
{{ Specifically, I suggest a paradigm shift where the one equation used over the last 40 years of $E_{\rm bind} =E_{\alpha_{\rm CE}} $,
be replaced by
\begin{equation}
 E_{\rm bind} = E_{\rm jets} +E_{\alpha_{\rm CE}},
\label{eq:jetsbind}
\end{equation}
where in many cases, most cases of surviving binary systems, $ E_{\rm jets} > E_{\alpha_{\rm CE}}$.
Here $E_{\rm bind}$ is the binding energy of the envelope,  $E_{\alpha_{\rm CE}}$ is the energy released by the binary
gravitational energy in the $\alpha_{\rm CE}$-prescription (eq. \ref{eq:alphace1}), and $E_{\rm jets}$ is the energy carried by the
jets launched by the companion (eq. \ref{eq:ejet1}).
One implication of he proposed paradigm shift is that the numerical study of the common envelope evolution can substantially move forward only if jets launched by the secondary
star are incorporated into CE simulations. }}

The different phases when jets might be launched from an accretion disk around a main sequence (MS) star are summarized in Table 1 and Fig. \ref{figure:fig1}.
The relevant phase is phase c, when the secondary is in the inner part of the CE and the envelope is highly oblate (flat).
{{{ {The formation of an accretion disk around the companion and the launching of energetic jets during phase c is the main new ingredient introduced in the present study.} }}}
In phase c, I suggest, accretion from material residing outside the secondary orbit, a circumbinary thick disk or oblate envelope, ensures sufficiently high
specific angular momentum to form an accretion disk.
The typical amount of mass required to be accreted onto a MS star to eject the envelope is given by equation (\ref{eq:macc1}).

Due to the orbital velocity of the secondary (eq. \ref{eq:vorb1}), the opening angle of the jets over an orbit will be large, and the jets will interact with a large
volume of the CE along the polar directions.
Typical jets velocities of $v_{\rm jet} \ga 500 \km \s^{-1}$ imply that the shocked jets' material form hot low density bubbles.
These bubbles then remove a large fraction of the envelope mass. After the removal of the CE, leftover circumbinary gas is likely to survive. Its life can be
prolong with fall back nebular gas, leading to the formation of post-CE jets (PCEJs).
The presence of PCEJs has been deduced from observations of some PNe (e.g., \citealt{Huggins2007, Tocknelletal2014}).
The PCEJs can form hot bubbles in the descendant elliptical planetary nebula (PNe) and be observed as diffuse X-ray emission \citep{Akashietal2008}.
The diffuse X-ray emission observed in some elliptical PNe might hint at the operation of the jet feedback mechanism (JFM) in the final removal of
the CE \citep{Freemanetal2014}.

Let me end by listing the basic ingredients of the proposed scenario for a final CE removal by a JFM.
\begin{enumerate}
\item The gravitational energy released by the in-spiraling core-secondary system is essential in inflating the envelope and reducing its binding energy.
\item The angular momentum transferred from the core-secondary system is essential in forming a flatten envelope (highly oblate), which turns into
a thick circumbinary disk that feeds the accretion disk around the secondary star.
\item For these two process to operate, the orbital separation must be small, $\sim 5-30 R_\sun$ for an AGB star with a MS companion.
That is, the JFM starts to operate only when the secondary is deep in the envelope.
\item The orbital separation continues to decrease mainly due to tidal interaction
with the thick circumbinary disk (migration).
\item A large fraction of the mass in the accretion disk, $\sim 20-40\%$, is launched in two opposite jets.
The accreted mass amounts to $\sim 0.1$ times the core mass, and the jets are launched with typical velocities of $\sim 700 \km \s^{-1}$ when the
companion is a MS star.
\item Typically, the jets' material is shocked to temperatures above those of the envelope, and hot bubbles are formed. These bubbles
interact with the envelope, much as X-ray deficient bubbles interact with the intracluster medium of clusters of galaxy \citep{Sokeretal2013},
and remove part of the envelope.
\item The very inner part of the giant envelope need not be expelled, as it can be accreted by the companion as part of the JFM. This
reduces the energy required to supply to the ejected envelope.
\item The removal of the envelope is composed of a negative-feedback part where the removal of the envelope reduces accretion rate,
and a positive-feedback part where the inflated bubbles push CE gas toward the equatorial plane and resupply material to the
circumbinary disk.
\item In many cases the jets will be active well after the ejection of the CE. These jets will further shape the descendant PNe and
might form extended X-ray emission in young elliptical PNe.
\item As the jets are not expected to be effective in all cases, in many cases the CE process ends in merger.
\end{enumerate}

{{{I thank two anonymous referees for comments that brought me to improve this manuscript. }}}
This research was supported by the Asher Fund for Space Research at the Technion.

\end{document}